\documentclass[preprint,11pt]{aastex}

\usepackage{color}

\shorttitle{Mn I lines in the H-band}
\shortauthors{Andersson, Ryde and Grumer {\it et al.}}

\begin{document}

\title{Hyperfine dependent $gf$-values of Mn I lines in the 1.49 -- 1.80 $\mu$m H-band}

\author{
M.~Andersson\altaffilmark{1,2},
J.~Grumer\altaffilmark{3}, 
N.~Ryde\altaffilmark{4}, 
R.~Blackwell-Whitehead\altaffilmark{4}, 
R.~Hutton\altaffilmark{1,2,6}, 
Y.~Zou\altaffilmark{1,2},
P.~J\"onsson\altaffilmark{5} 
and T.~Brage\altaffilmark{3}}

\altaffiltext{1}{The Key lab of Applied Ion Beam Physics, Ministry of Education, China}
\altaffiltext{2}{Shanghai EBIT laboratory, Modern physics institute, Fudan University, Shanghai, China}
\altaffiltext{3}{Division of Mathematical Physics, Department of Physics, Lund University, Sweden}
\altaffiltext{4}{Department of Astronomy, Lund University, Sweden}
\altaffiltext{5}{School of Technology, Malm\"o University, Sweden}
\altaffiltext{6}{Author to whom any correspondence should be addressed (e-mail: \textit{rhutton@fudan.edu.cn})}

\begin{abstract}
The three Mn I lines at 17325, 17339 and 17349 \AA~are among the 25 strongest lines ($\log(gf)>0.5$) in the H-band. They are all heavily
broadened due to hyperfine structure and the profiles of these
lines have so far not been understood. Earlier studies of these
lines even suggested that they were blended.
In this work, the profiles of these three infra-red (IR) lines have been studied
theoretically and compared to experimental spectra to assist in the
complete understanding of the solar spectrum in the IR. It is shown that
the structure of these lines can not be described in the conventional
way by the diagonal $A$ and $B$ hyperfine interaction constants.
The off-diagonal hyperfine interaction not only has large impact on the
energies of the hyperfine levels, but also introduces a large intensity
redistribution among the hyperfine lines, changing the line profiles
dramatically. By performing large-scale calculations of the diagonal and
off-diagonal hyperfine interaction and $gf$-values between the upper
and lower hyperfine levels and using a semi-empirical fitting procedure,
agreement between our synthetic and experimental spectra was achieved.
Furthermore, we compare our results with observations of
stellar spectra. The spectra of the Sun and the K1.5 III red giant
star Arcturus were modelled in the relevant region, $1.73-1.74$ $\mu$m
using our theoretically predicted $gf$-values and energies for each
individual hyperfine line. Satisfactory fits were obtained and
clear improvements were found using our new data compared with the old
available Mn I data. A complete list of energies and $gf$-values for all the 
$3d^54s({^7S})4d$~e$^{6}$D -- $3d^54s({^7S})4f$~w$^{6}$F
hyperfine lines are available as supplementary material online,
whereas only the stronger lines are presented and discussed in
detail in this paper.
\end{abstract}


\section{Introduction}\label{intro} 
High wavelength resolution spectrographs on satellite borne and
ground based telescopes can resolve many of the absorption features
in stellar spectra. In particular, line broadening effects such as
hyperfine structure (HFS) and isotope shifts can be observed in the
spectra of the Sun (\cite{Livingston91}) and other stars (e.g.
Arcturus, \cite{arcturusatlas}). HFS increases the line width and
decreases the peak intensity of the line profile. Omitting such effects
will introduce an error in the measurement of the wavelength and the derived abundance
from a stellar spectrum (\cite{Prochaska00}). It was shown by \cite{Jomaron99},
that if the HFS is not taken into account, the abundance of manganese in HgMn stars
can be overestimated by up to three orders of magnitude. It was also
shown that if the HFS of manganese is included as a crude estimate, the derived
abundance may still be up to 4 times too large. \cite{Jomaron99} suggested that
the lack of accurate information about the HFS structure is the
largest contributing factor to the uncertainty when estimating the abundance of manganese.

Hyperfine splitting of \ion{Mn}{1} line profiles can be observed in
the visible (\cite{Abt52}) and infrared (\cite{Swensson66}) spectrum of
the Sun. In particular, \cite{Melendez99} used the hyperfine
split solar line profiles of \ion{Mn}{1} in the H--band (1.49 to
1.80 $\mu$m) and J--band (1.00 to 1.34$\mu$m) to measure the
wavelength of the hyperfine component lines for the study of
metal--rich stars in the Galactic bulge. However, \cite{Melendez99}
notes that the strong (log({\it gf}) $> -0.5$) \ion{Mn}{1} lines at 17325,
17339 and 17349~\AA\ have broad hyperfine splitting and appear to be
blended with unknown features in the solar spectrum. Fitting these
profiles is particularly difficult because there are no hyperfine
structure constants in the literature for the upper levels
($3d^54s({^7S})4f$~w$^{6}$F$_{J}$) of these transitions. Furthermore, the fine
structure energy level values for w$^{6}$F$_{J}$ are poorly known
and the NIST atomic level database \citep{Ralchenko08} provides values
from the work of \cite{Catalan64} who used the Land\'{e} interval
rule to calculate the energy level values from blended
transitions in the UV. The high uncertainty in the energy level
values for w$^{6}$F$_{J}$ increases the uncertainty in the line
identification and thus the fitting of the blended infra-red (IR) transitions.
The laboratory line list for \ion{Mn}{1} transitions in the IR by
\cite{Taklif90} includes wavelengths for the 17325, 17339 and
17349~\AA\ lines but \cite{Melendez99} misinterprets the \cite{Taklif90}
line list and identifies the 17325~\AA\ feature as only the
$3d^54s({^7S})4d~e^{6}$D$_{9/2}$ -- $3d^54s({^7S})4f$~w$^{6}$F$_{9/2}$
transition. However, \cite{Taklif90} identifies the 17325~\AA\ line as a
blend of three transitions e$^{6}$D$_{9/2}$ -- w$^{6}$F$_{9/2}$,
e$^{6}$D$_{9/2}$ --  w$^{6}$F$_{7/2}$ and e$^{6}$D$_{9/2}$ -- w$^{6}$F$_{11/2}$.
A similar misinterpretation is given by \cite{Melendez99} for the 17339 and
17349~\AA\ features, which are reported by \cite{Taklif90} to be the
following blended lines; e$^{6}$D$_{7/2}$ -- w$^{6}$F$_{9/2, 7/2,
5/2}$ (17339~\AA ) and e$^{6}$D$_{5/2}$ -- w$^{6}$F$_{7/2, 5/2,
3/2}$ (17349~\AA ).

A high resolution study of the spectrum of neutral manganese from
the IR to the vacuum UV was reported in the thesis by
\cite{Blackwell03} and further studies of the hyperfine structure are
given in \cite{Blackwell05}. It was noted in
\cite{Blackwell03} that it was not possible to fit the transitions
from the w$^{6}$F term using diagonal hyperfine interaction constants.
In this work Blackwell-Whitehead also indicated that the hyperfine splitting of the
blended 17325, 17339 and 17349~\AA\ lines may require a more
detailed theoretical analysis to fully understand these line
profiles. Furthermore, given that 17325, 17339 and 17349~\AA\ lines
are within the 25 strongest (log({\it gf}) $> -0.5$) \ion{Mn}{1} lines in
the H-band, a study of their profiles will assist in the complete
interpretation of the solar spectrum in the IR.

We show that the 17325, 17339 and 17349~\AA\ line features are hyperfine
split, blended features of the e$^{6}$D$_{9/2}$ -- w$^{6}$F,
e$^{6}$D$_{7/2}$ -- w$^{6}$F and e$^{6}$D$_{5/2}$ -- w$^{6}$F transitions.
These features can not be described using diagonal hyperfine interaction
constants due to a strong off-diagonal hyperfine interaction which in
some cases leads to such a large mixing between the hyperfine levels
that the $J$-quantum number looses its meaning. Thus to assist in the
analysis of these transitions in stellar spectra we provide individual, relative
line positions and {\it gf}-values for each hyperfine transition.
Furthermore, we suggest that the solar line profiles for 17325, 17339 and
17349~\AA\ (air wavelengths) can be explained by off-diagonal hyperfine interaction
and we claim that these lines are not significantly affected by
unknown blends in the solar spectrum. These lines should therefore be useful in the 
analysis of stellar spectra, for instance in the determination of stellar Mn abundances.

\section{Hyperfine interaction}
In isotopes with a non-zero nuclear spin, $I$, there is an interaction
between the electromagnetic moments of the nucleus and the electrons, which
is often referred to as the hyperfine interaction. This interaction couples the
total electronic angular momentum $J$ and the nuclear spin $I$, to a new total
angular momentum $F$. As a consequence of this interaction, each fine structure
level is split up into several closely spaced hyperfine levels. If the energy
separations between the fine structure levels are large compared to the separations 
due to the hyperfine interaction, the energies of the hyperfine levels can be
calculated using lowest order perturbation theory
\begin{eqnarray}\label{HPF-diag}
\lefteqn{ E_{hpf}(\gamma JF) = E_{fs}(\gamma J) + \frac{1}{2}AK } \\
 & & \quad + B\frac{(3/4)K(K+1)-J(J+1)I(I+1)}{2I(2I-1)J(2J-1)}, \nonumber
\end{eqnarray}
where $E_{fs}(\gamma J)$ is the energy of the fine structure level,
\begin{equation}
K=F(F+1)-J(J+1)-I(I+1),
\end{equation}
and $A$ and $B$ are the hyperfine interaction constants. The label $\gamma$ denotes 
the quantum numbers required to identify the fine structure level.

Manganese has only one stable isotope with a nuclear spin of $I=5/2$
and a strong nuclear magnetic dipole moment, $\mu_I=3.4687$ nuclear
magnetons, as well as a small electric quadrupole moment, $Q=0.32$
barns (\cite{Lide2003}), leading to a potentially strong hyperfine
interaction. The open $4s$ shell in the $3d^54s(^7S)4f~w{^6}$F term
gives rise to a strong hyperfine interaction. At the same time
the fine structure of this term is very small, resulting in a strong
off-diagonal hyperfine interaction, i.e. interaction between
hyperfine levels derived from different fine structure levels described
in the diagonal approximation using the $A$ and $B$ interaction
constants.

In order to describe a system where the fine and hyperfine structure energy
splitting is of the same order of magnitude one has to use higher orders of
perturbation calculation or use a matrix formalism to take the off-diagonal
hyperfine interaction into account. In this work we have used the latter
approach. We will not describe the method in detail in this paper
but refer to three earlier papers which were based on similar
approaches, \cite{AnderssonGa4f}, \cite{Grumer2010} and \cite{Andersson2012}.

To describe the hyperfine states we couple the {\it J}-dependent electronic wave 
function $|\gamma J\rangle$ to the nuclear wave function $|I\rangle$ using standard
coupling theory to build $F$-dependent wave functions $|\gamma JIF\rangle$ (FSF)
which form a set of basis functions in our calculation. The Atomic State
Function (ASF), $|\Gamma F\rangle$, representing the hyperfine mixed hyperfine
levels, is written as a linear combination of the FSFs as
\begin{equation}\label{HPF-wfn}
|\Gamma F\rangle = \sum_i c_i |\gamma_i J_i IF\rangle ,
\end{equation}
where $c_i$ are expansion coefficients.

In the calculation of the lower hyperfine levels all possible
$|e^6D~JIF\rangle$ and $|e^8D~JIF\rangle$ FSFs were used and for the upper
hyperfine levels all $|w^6F~JIF\rangle$ and $|w^8F~JIF\rangle$. Using
these basis functions the hyperfine interaction Hamilton matrix was
set up and diagonalized to yield hyperfine level energies and the
expansion coefficients of the ASFs.

The transition operator acts only on the electronic part of the
wave function and the nuclear part can therefore be decoupled. The {$gf$}-values
of the hyperfine transitions can then be calculated in terms of $J$-dependent
transition matrix elements as
\begin{eqnarray}\label{eq:tot-gf}
gf & = & \frac{8\pi^2m_eca_0^2\sigma}{3h} (2F_i+1)(2F_j+1) \nonumber
\\
 & & \times \Bigg| \sum_{i} \sum_{j} (-1)^{J_i}c_{i}
c_{j} \nonumber
\\ &   & \times \left\{
  \begin{array}{ccc}
        F_i & J_i & I \\
        J_j & F_j & 1
  \end{array}
\right\} \langle \gamma_i J_i \| {\bf D}^{(1)} \| \gamma_j J_j\rangle
\Bigg|^2 .
\end{eqnarray}
For details, see \cite{Grumer2010}.

\section{Method of Calculation}
The calculations were based on first optimizing wave functions for
the lower e$^6$D$_J$ and the upper w$^6$F$_J$ fine structure levels using
the relativistic atomic structure package \textsc{Grasp2k}
\citep{Jonsson2007}. These programs are based on the multiconfiguration
Dirac-Hartree-Fock (MCDHF) method as outlined by \cite{Grant2007}. The
even and the odd states were optimized in two separate calculations.
Using the electronic wave functions the Hamiltonian matrix, including the
hyperfine interaction, was constructed and diagonalized using the 
\textsc{Hfszeeman} program (\cite{AnderssonHFSZEEMAN}) to give the hyperfine level 
energies as well as the corresponding wave functions in the form of 
equation~(\ref{HPF-wfn}).

Having the ASFs for the hyperfine levels, the $gf$-values of the
transitions were calculated using a newly developed code, 
connected to the \textsc{Grasp2k} suite of programs, which determines rates of $F$-
dependent transitions in a general manner (\cite{GrumerHfstrans}). Note that this 
code also allows for an external magnetic field in cases when it is large enough to 
be non-negligible. The program is based on equation~(\ref{eq:tot-gf}) and 
uses the mixing coefficients from \textsc{Hfszeeman} together with the {\it J}-
dependent transition matrix elements from a slightly modified version of the 
\textsc{Grasp2k} transition program.

To investigate the importance of the off-diagonal hyperfine
interaction, two different calculations were performed. The first we
have named the {\it Complete} calculation and the second the {\it
Diagonal} calculation. In the {\it Complete} calculation, the full
hyperfine interaction matrix was used, whereas in the {\it Diagonal} calculation 
only the diagonal hyperfine interaction matrix elements were included.
The {\it Diagonal} calculation therefore corresponds to describing
the hyperfine interaction in terms of {\sl A} and {\sl B} hyperfine
interaction constants.

\section{Laboratory Measurements}
The emission spectrum of manganese, fig 2 to 4, was recorded at the
National Institute of Standards and Technology (NIST) with the NIST
2-m Fourier transform spectrometer \citep{Nave1997} using resolutions
of 0.008 to 0.03 cm$^{-1}$, which is sufficient to fully resolve the
Doppler broadened line profiles of the transitions. The light-source
used was a water-cooled hollow cathode lamp \citep{Blackwell03,Blackwell05}.
Owing to the brittle nature of pure manganese, the cathodes were made
of an alloy of 95$\%$ Mn and 5$\%$ Cu. The hollow cathode was run at a
current of 1.5 A, with 1.9 Torr of Ne as a buffer gas.

\section{Synthetic spectra}
The laboratory spectra were recorded using a
hollow cathode, and the line intensities of a spectrum from such a
light source should be proportional to the {$gf$}-values under the
assumption that the transition rates are much higher than the collision rates
and that line intensities from the same multiplet are compared. The synthetic 
spectra were generated by giving each hyperfine line a Voigt profile and 
the Full Width Half Maximum (FWHM) was fitted to the experimental spectrum.

To be able to reproduce the experimental spectrum it was necessary
to adjust our {\it ab initio} energies. The energies of the e${^6}$D
levels have been determined experimentally but the fine structure of
the w${^6}$F term is poorly known. The NIST atomic level database
\citep{Ralchenko08} provides values from the work of \cite{Catalan64}
who used the Land\'{e} interval rule to calculate the energy level
values from blended transitions in the UV. However, since the hyperfine
interaction is of the same order as the fine structure splitting, this
method should be considered invalid.

To start from accurate fine structure energies is of great importance
since the hyperfine mixing is very sensitive to the fine structure
splitting. The fine structure splitting of the w${^6}$F levels is very
small compared to the term splitting between w${^6}$F and 
$3d^54s({^7S})4f$~z${^8}$F. Since the fine structure is associated
with the spin-orbit interaction and this interaction is responsible
for the mixing between the levels of w${^6}$F and z${^8}$F, the
fine structure of w${^6}$F should be close to the Land\'e interval
rule,
\begin{equation}
E_{fs}(LSJ) = \frac{{\cal C}(LS)}{2}[J(J+1)-L(L+1)-S(S+1)]
\end{equation}
where ${\cal C}$ is the Land\'e interval constant. We used this
argument as a start when trying to reproduce the experimental
spectrum. By using the error defined by the least square fit between
the synthetic and experimental spectra, the Land\'e interval
constant was varied to find the best fit. The resulting value of ${\cal C}$ was in 
this case found to be $0.0150$ cm$^{-1}$. We will refer to results using 
this method as {\it Land{\'e} Fitted}.

To further improve the fitting and allow for deviations from the
Land\'e interval rule, the energies of the independent fine structure
levels were varied. This was also done by using the error defined by
the least square fit between the synthetic and experimental spectra.
Using this approach good agreement was found between the synthetic
and experimental spectra. Results based on this model will be referred to
as {\it Level Fitted}.

In general, even if theoretically predicted $gf$-values are close to
experimental ones, the predicted energies for the lines are not of experimental
accuracy. To improve our synthetic spectra we therefore made a 
final adjustment where we allowed for small variations of the hyperfine level 
energies when fitting to experimental spectra. In these adjustments, only
the hyperfine level energies were changed, whereas all {$gf$}-values were kept 
fixed. We again used the error defined by the least square fit between the
synthetic and experimental spectra to find a better fit. We allowed
for variations for both the lower and upper hyperfine levels
resulting in 54 free parameters. Our computer power was not large
enough to handle so many parameters, but we had to try to improve
the spectra stepwise, going from the upper to the lower end of the
spectrum. Using this approach we could reproduce the experimental
spectrum to very high accuracy. We will refer to results including this final adjustment as {\it Hyperfine Adjusted}.

\section{Results and Discussion \label{sec:results}}
Besides trying to reproduce the experimental spectra and obtaining information
about all individual hyperfine lines, we also investigated the importance
of the off-diagonal hyperfine interaction, and how important the different
steps of our fitting procedure were to reproduce the experimental spectra.
The influence of the off-diagonal hyperfine interaction can be found by comparing
the results from {\it Diagonal} and {\it Complete} calculations (see section 3).

Using the {\it Land{\'e} Fitted} method, described in section 5, we performed {\it 
Diagonal} and {\it Complete} calculations trying to fit the synthetic spectra to
experiment by varying the Land\'e interval constant. The former of these
we will refer to as the {\it Diagonal} and the latter as {\it Complete Lande Fitted}
({\it CLaF}) calculation. Comparing the spectra from these two calculations, the 
importance of the off-diagonal hyperfine interaction can be found.

Including the off-diagonal hyperfine interaction we performed two further 
calculations, the {\it Complete Level Fitted} ({\it CLeF}) using the {\it Level 
Fitted} procedure described in section 5, and the {\it Complete Hyperfine Adjusted} 
({\it CHA}) using the {\it Hyperfine Level Adjusted} method also described in section 5.

Comparing the synthetic spectra from the {\it CLaF} calculation to the one from
the {\it CLeF} calculation, the influence of adjusting the fine structure energies 
of w$^6$F from the Land\'e interval rule can be determined. Finally, the
impact of adjusting the individual hyperfine level energies to the
synthetic spectra can be investigated by comparing the spectra from the {\it CLeF}
calculation on the spectra from the {\it CHA} calculation.

\begin{figure}
\plotone{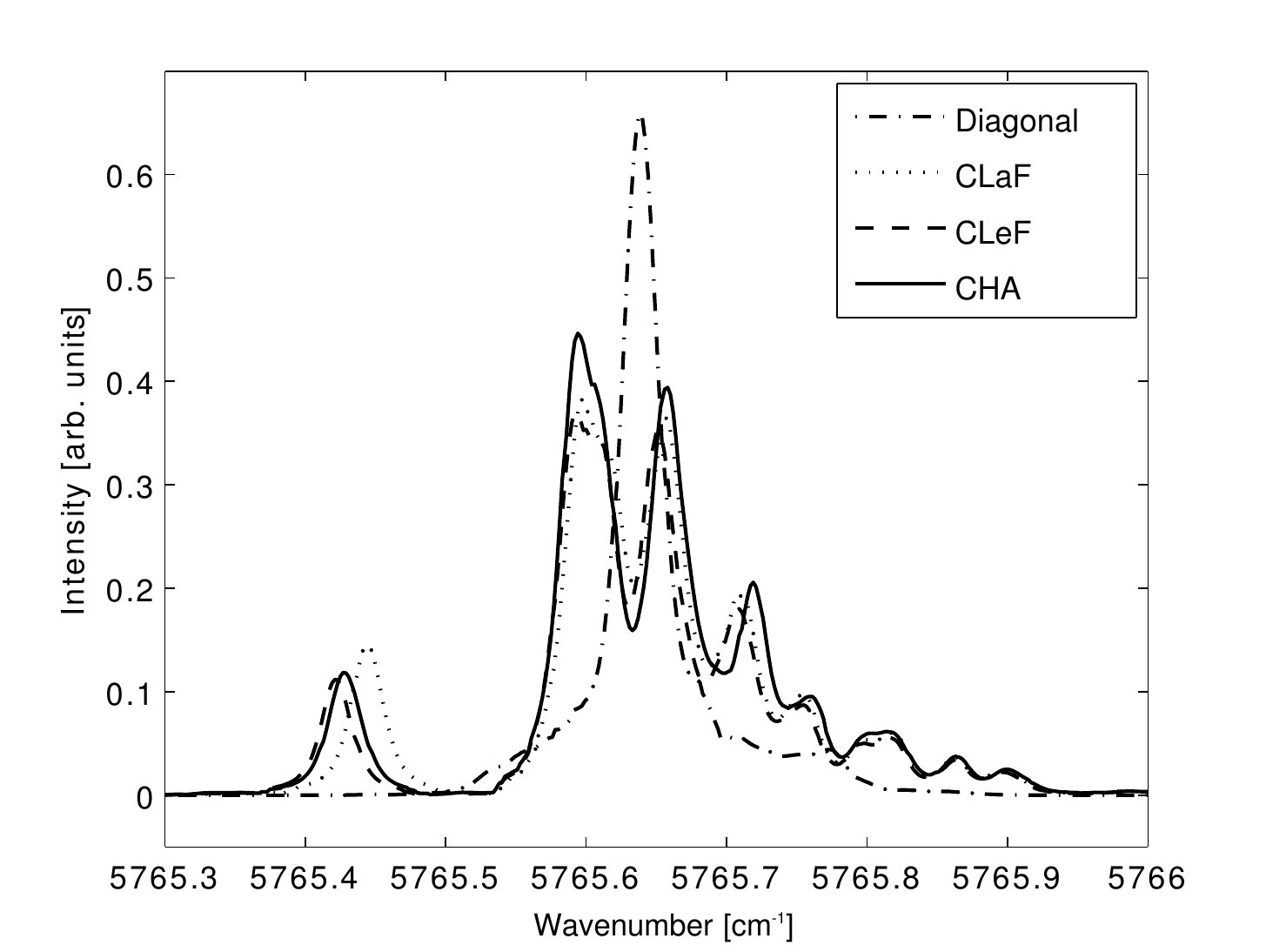}\caption{\label{Fig-DiffTypes6D35}Four
different synthetic spectra of the e$^6$D$_{7/2}$--w$^6$F
transitions (the 17339\AA~line). The dash-dotted spectrum is from the Diagonal calculation,
the dotted from the Complete Lande Fitted calculation(CLaF), the dashed from the Complete Level
Fitted (CLeF) and the solid from the Complete Hyperfine Level Adjusted (CHA).}
\end{figure}

To see how the synthetic spectra changed through the {\it Diagonal}, {\it CLaF},
{\it CLeF} and {\it CHA} calculations, we have chosen to present the 17339 \AA 
~spectral feature, corresponding to the e$^6$D$_{7/2}$--w$^6$F hyperfine lines, for 
these four calculations in Figure~\ref{Fig-DiffTypes6D35}. The differences between 
the four different synthetic spectra for the 17325, 17349, 17357 and 17362 \AA~
spectral features, corresponding to the  e$^6$D$_{9/2,5/2,3/2,1/2}$~--~w$^6$F
hyperfine lines, follows much the same pattern.

From Figure~\ref{Fig-DiffTypes6D35} it is found that the {\it
Diagonal} calculation in principle predicts one strong peak
surrounded on both sides with some weak structure. The spectrum from
the {\it CLaF} calculation predicts a much wider and more complex
structure and there are in principle no similarities between the two
spectra. It should be pointed out that these differences are
entirely due to the off-diagonal hyperfine interaction, which not
only affects the energies of the hyperfine levels but also has a
very large impact on the {$gf$}-values of the individual hyperfine
transitions. It is clear from this picture that the hyperfine
levels of w$^6$F can not be described in terms of {\sl A} and {\sl B}
hyperfine constants.

Moving from the {\it CLaF} to the {\it CLeF} spectrum it is found
that the position and the intensities of some lines have been
slightly changed, but the differences are rather small. The same is
found going from {\it CLeF} to {\it CHA}. It should be pointed out
that going from {\it CLaF} to {\it CLeF} did not only
change the position of the hyperfine lines but also slightly changed
the {$gf$}-values, whereas going to {\it CHA} only changed the position of the
hyperfine lines whereas the {$gf$}-values were the same as for {\it
CLeF}.

Below we give some results in detail for the three spectral regions of
e$^6$D--w$^6$F of greatest astrophysical interest. For each group of peaks
we present a figure (Figure~2-4) where we have plotted our synthetic
spectrum from the {\it CHA} calculation compared to the experimental one.
In each figure we also present the synthetic spectrum from the
{\it Diagonal} calculation as an inset plot.

The off-diagonal hyperfine interaction gives rise to many new transitions
and the total number of transitions within each sub-spectrum can
therefore be very large, and the total transition list is therefore
too long to be published in the paper version of this article
but is available as online supplementary material. Instead we have chosen to
present those lines that have a {$gf$}-value equal to or greater than 10\%
of the largest {$gf$}-value within each sub-spectrum. These lines give a
good description of the spectrum and the additional lines only make
small changes and the reduced list is therefore sufficient for
discussing the results.

In each table, various information about each hyperfine line is presented. In
the first column the $F$-value of the w$^6$F hyperfine level is given. The second
column is labelled HFS, and refers to HyperFine State. This is an index identifying
the different hyperfine levels in the calculations. Since the off-diagonal hyperfine 
interaction introduces a large mixing between the hyperfine 
states in this system, the $J$-value is no longer a good quantum number and can 
therefore not be used to identify the hyperfine levels. Instead we gave each 
hyperfine level an identification number according to the energy order of the 
hyperfine levels within each parity symmetry. In column three and four the 
corresponding information is given for e$^6$D hyperfine levels. In column five, the 
wavenumber from the {\it CHA} calculation for the hyperfine transition is given,  
and in the sixth column the gf value from the same calculation is given.
In column seven the $gf$-value from the corresponding hyperfine transition in the 
{\it Diagonal} calculation is given and in the last column is the difference in 
$gf$-value of the {\it Diagonal} calculation relative to the {\it CHA} calculation.
The complete e$^6D$~--~w$^6F$ line list can be found as supplementary
material online.

The accuracy of the wavenumbers should undoubtedly be high as the atomic energy 
structure is deduced from high quality wavefunctions and subsequently anchored to 
high precision laboratory spectra. The values in the tables are therefore given with 
four decimals. The estimated uncertainties of the relative positions, which are the 
important quantities here, are smaller than $\pm 0.02$ cm$^{-1}$. This is better 
than what is required for stellar spectroscopy. It should be made clear that 
uncertainties of the absolute wavenumbers are slightly higher as they are dependent 
on the calibration of the experimental spectra. 

\subsection{The 17325 \AA~line}
\begin{figure}
\plotone{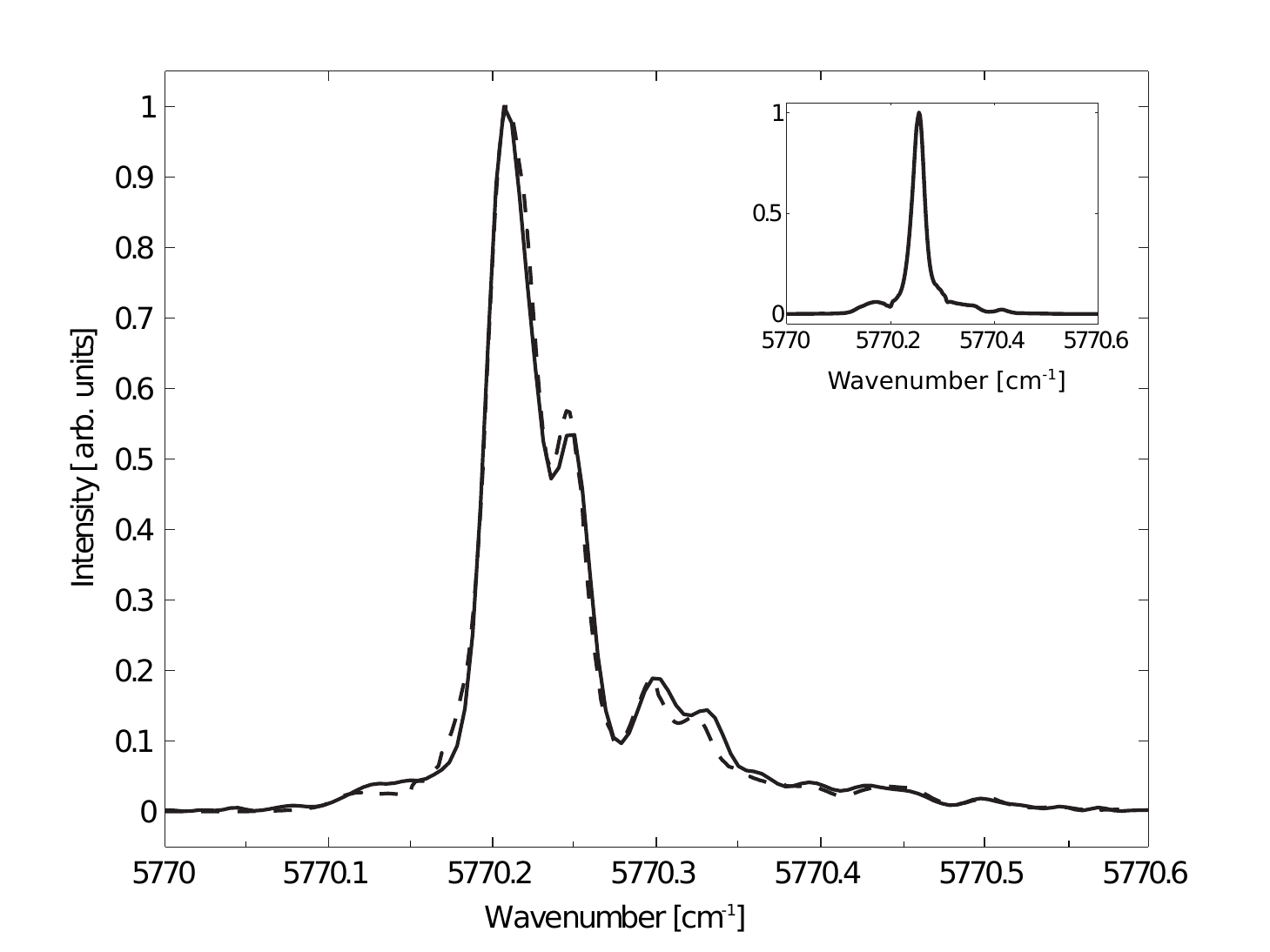}\caption{\label{Fig-6D45} The experimental
spectrum (solid line) compared to the {\it CHA} spectrum (dashed line)
for the e$^6$D$_{9/2}$--w$^6$F transitions (the 17325\AA~line). The
{\it Diagonal} spectrum is presented as an inset plot.}
\end{figure}

We start by investigating the 17325 \AA~line corresponding to the e$^6$D$_{9/2}$--
w$^6$F transitions in the interval $5770.0-5770.6$ cm$^{-1}$. The result is 
presented as a plot in Figure~\ref{Fig-6D45} and in detail in Table~\ref{Tab-6D45}. 
From Figure~\ref{Fig-6D45} it is found that our {\it CHA} synthetic spectrum 
reproduces the features of the experimental one, whereas the
{\it Diagonal} calculation predicts a structure that is too simple. The main
differences between the {\it Diagonal} and the {\it CLeF} spectra are
the two peaks emerging at $5770.30$ cm$^{-1}$ and $5770.33$ cm$^{-1}$.
From Table~\ref{Tab-6D45} it is found that the predicted {$gf$}-values for the
hyperfine transitions making up these lines are about 25\% smaller in the {\it 
Diagonal} calculation than in the {\it CLeF} calculation. The fact that these two 
lines emerge in the spectrum generated from the {\it CLeF} calculation is 
only partly explained by the enhancement of the $gf$-values. The main underlying 
reason is rather the shift in energy induced by the off-diagonal hyperfine 
interaction.

\subsection{The 17339 \AA~line\label{sec:line17339}}

The 17339 \AA~line is situated in the region $5765.3 - 5766.0$ cm$^{-1}$.
The comparison between the experimental and the {\it CHA} spectra is
presented in Figure~\ref{Fig-6D35}. In the same figure the
spectrum from the {\it Diagonal} calculation is also included as an inset plot.
Starting with the {\it Diagonal} spectrum it is found that it in principle predicts 
only one peak, whereas the experimental spectrum of the e$^6$D$_{7/2}$--w$^6$F
transitions is much more complex. The {\it CHA} synthetic spectrum on the other
hand reproduces all features of the experimental one. The largest difference between
{\it CHA} and experiment is the line at $5765.43$ cm$^{-1}$. In {\it CHA} this
peak is approximately 10\% lower than experiment and it is positioned at an energy
that is $0.007$ cm$^{-1}$ too low.

The results are presented in detail in Table~\ref{Tab-6D35}. It is
found that there are much larger differences between the {\it
Diagonal} and {\it CLeF} calculation for this part of the spectrum than
for the 17325 \AA~spectral feature. Even the {$gf$}-values for the strongest and 
second strongest lines differ by 10\% and 28\% respectively. Even more notable is 
that there are three lines in the list that have {$gf$}-values which are identically
zero in the {\it Diagonal} calculation, and that the strongest of these have a 
{$gf$}-value which is 14\% of the strongest of all lines in this
part of the spectrum. It is clear from this list that the changes to the {$gf$}-
values due to the off-diagonal hyperfine interaction have a very large impact on the 
spectrum.

\subsection{The 17349 \AA~line}
The 17349 \AA~line is situated in the region $5761.7 - 5762.6$ cm$^{-1}$.
In the main plot of Figure~\ref{Fig-6D25} we compare our {\it CHA} synthetic 
spectrum to the experimental. The corresponding spectrum generated from the {\it 
Diagonal} calculation is presented as an inset plot. Comparing the {\it Diagonal} 
synthetic spectrum to experiment it is found that it is far off the target and that 
there are not many similarities between the two spectra for the e$^6$D$_{5/2}$--
w$^6$F transitions. However, there is a good resemblance between the {\it CHA} 
synthetic spectrum and the experimental one.

Inspecting the results of the e$^6$D$_{5/2}$--w$^6$F hyperfine transitions as 
presented in Table~\ref{Tab-6D25}, it is found that the differences between the {\it 
CLeF} and {\it Diagonal} calculation are even larger than for the 173325 and 
17339 \AA ~spectral features in the spectrum. The transition with the largest $gf$-
value differs by 36\% between the two calculations and the 4th strongest line, with 
a {$gf$}-value of 42\% of the largest, in the {\it CHA} calculation is a strictly 
forbidden transition in the {\it Diagonal} calculation. Actually, the 4th,
5th, 8th, 10th and 14th strongest lines of the 22 transitions in the line list are 
all induced by the off-diagonal hyperfine interaction and are absent in the {\it 
Diagonal} spectrum. This has of course a very large impact on the spectrum and is a 
further proof of the invalidity of describing the hyperfine structure of the
e$^6$D--w$^6$F spectrum in terms of A and B hyperfine constants.

\subsection{Uncertainties of the $gf$-values \label{sec:uncert}}
As it is hard to give any precise values of the \emph{absolute} $gf$-uncertainties 
in the present work, we focus the current discussion on the \emph{relative} 
uncertainties. These are undoubtedly also the most relevant quantities to discuss as 
the absolute values anyway are easy to rescale with an overall common factor, 
possibly evaluated from a comparison with a well-calibrated measurement. 
Nevertheless, even though there is little experience about hyperfine structure 
analyzes of a complexity comparable to the present work, we expect the overall 
uncertainty of the \emph{absolute} $gf$-values to be well below $10\%$.

\begin{figure}
\plotone{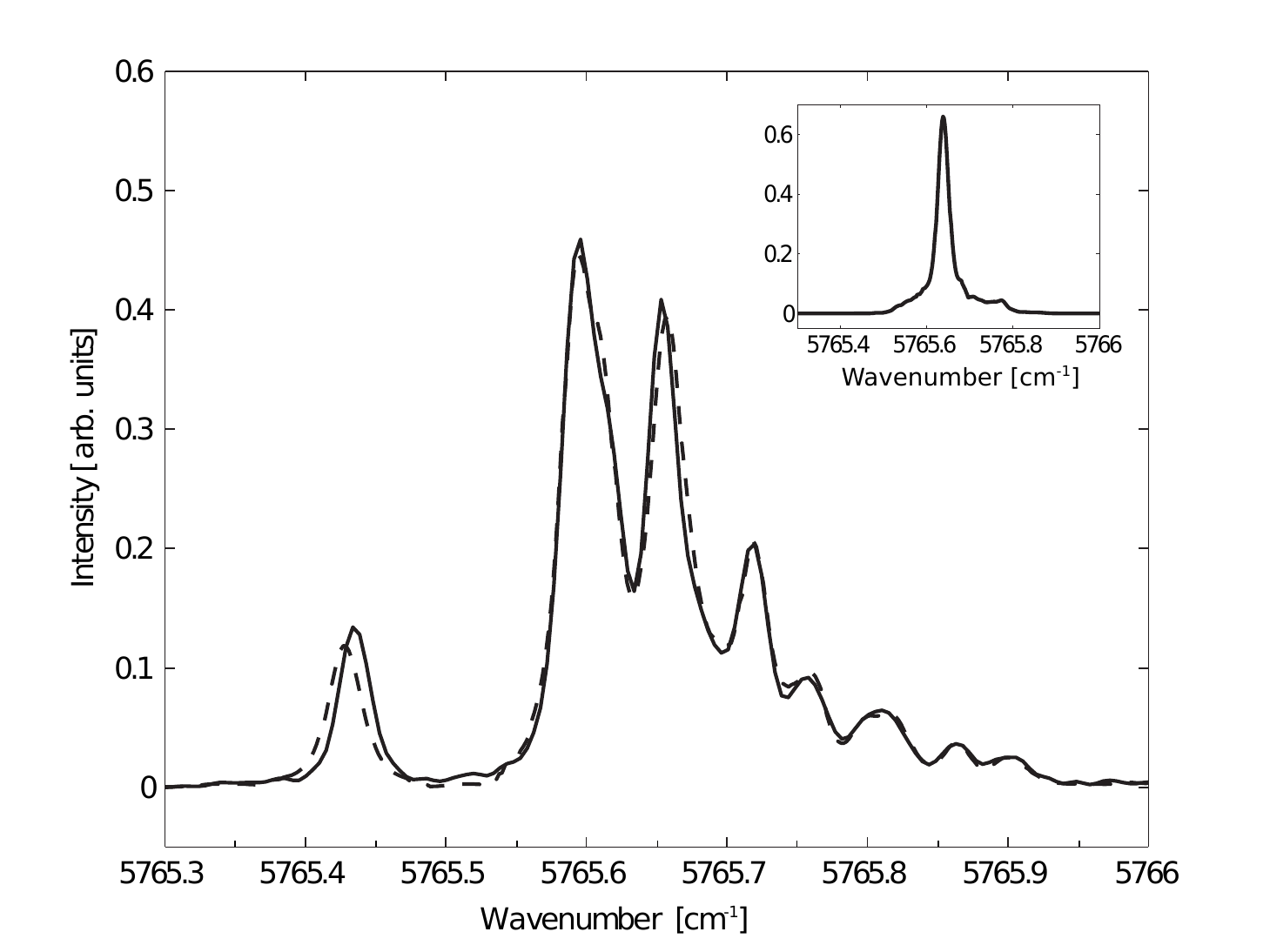}\caption{\label{Fig-6D35}The experimental
spectrum (solid line) compared to the {\it CHA} spectrum (dashed line)
for the e$^6$D$_{7/2}$--w$^6$F transitions (the 17339\AA~line). The
{\it Diagonal} spectrum is presented as an inset plot.}
\end{figure}

\begin{figure}
\plotone{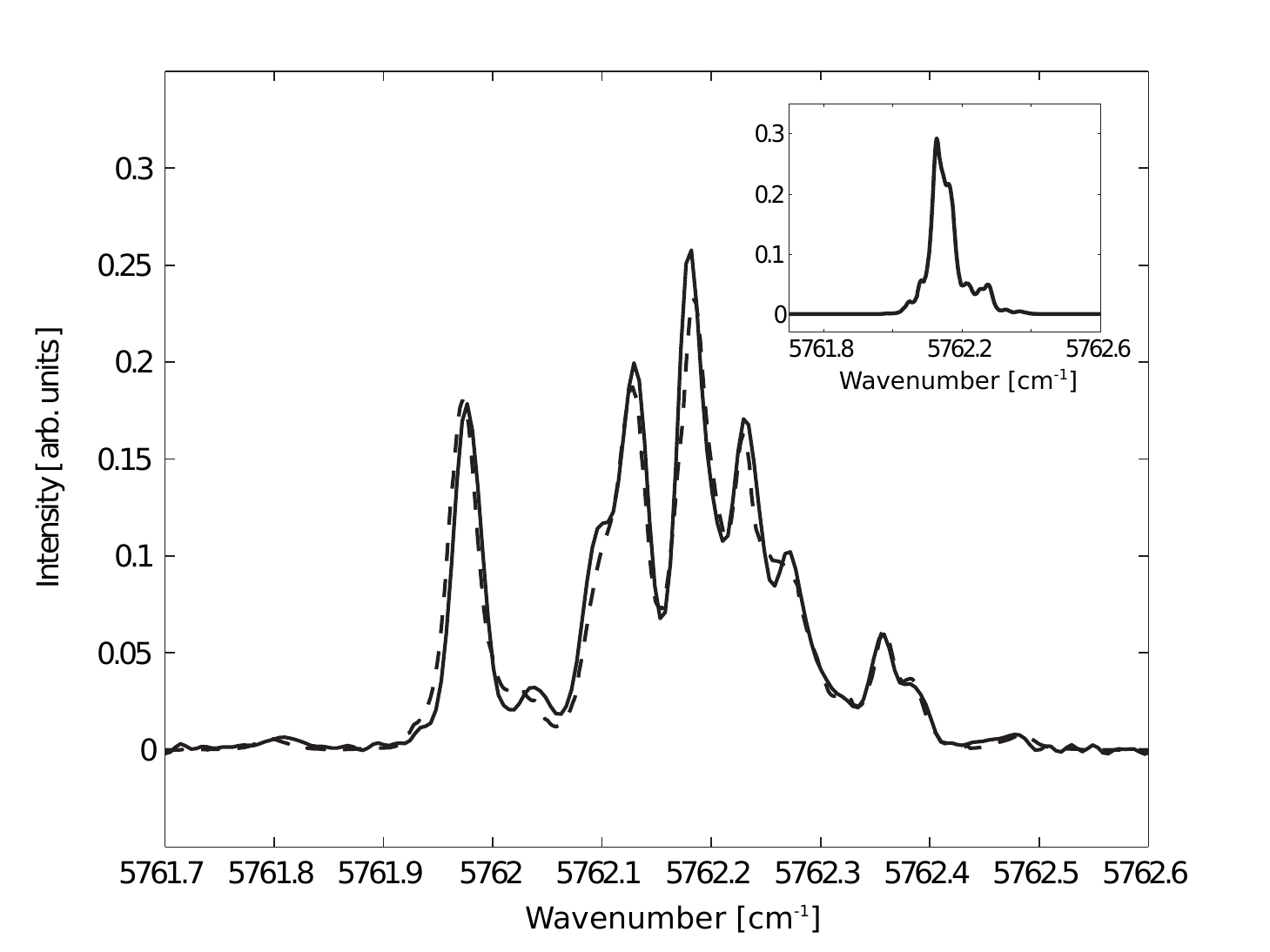}\caption{\label{Fig-6D25}The experimental
spectrum (solid line) compared to the {\it CHA} spectrum (dashed line)
for the e$^6$D$_{5/2}$--w$^6$F transitions (the 17349\AA~line). The
{\it Diagonal} spectrum is presented as an inset plot.}
\end{figure}

The uncertainty of the \emph{relative} $gf$-values can be estimated from comparisons 
of the synthetic spectra to the corresponding experimental spectra as presented in 
Figures \ref{Fig-6D45}, \ref{Fig-6D35} and \ref{Fig-6D25}. By investigating all 
hyperfine components of these spectra, one can conclude that the synthetic line 
which seems to fit worst with experiment is the left-most line of Figure 
\ref{Fig-6D35} at $5765.43$ cm$^{-1}$. This line has a $gf$-value which is about 
$10\%$ too small as compared to experiment which was noted above in Section 
\ref{sec:line17339}. It should however be clear that this is the worst case scenario 
as the synthetic spectrum could be scaled up with a factor to better fit this line 
with the spectrum and thereby instead overestimate the group of lines in the center 
of this part of the spectrum. One could therefore consider $10\%$ as an upper limit 
of the relative $gf$-uncertainty. Furthermore it should be noted that this line is 
an example of a transition which is not at all predicted by a conventional $A$ and 
$B$ hyperfine constant (or \emph{Diagonal}) analysis. Another line which doesn't fit 
perfectly with the experimental spectrum is the structure just right of the main 
peak in Figure \ref{Fig-6D45} at 5770.25 cm$^{-1}$. The $gf$-value of this line 
deviates from experiment by approximately $5\%$.
Apart from these two lines we judge the overall uncertainty of the \emph{relative} 
$gf$-values to be well within $5\%$.

\section{Modeling stellar Mn I lines}
A way to test our new calculations of the hyperfine splitting of the
Mn lines, is to compare with observations of stellar spectra. We
have therefore modelled the spectra of the Sun (of spectral type G2V)
and of the K1.5 III red giant star Arcturus ($\alpha$ Boo)  in the
relevant spectral region of $1.73-1.74\,\mu$m in order to be able to
compare with the atlases of these stars by \cite{Livingston91} and
\cite{arcturusatlas}, respectively.  The spectral resolution of
these atlases is sufficiently high to resolve the stellar spectral
lines. We calculated the synthetic spectra for atmospheres modelled
with  the {\sc marcs} code \citep{marcs:08}
\footnote{For the Sun we use $T_{\mathrm{eff}}=5770$ K, $\log
g=4.44$, $\xi_{\mathrm{micro}}=0.93$ km\,s$^{-1}$, and  solar
abundances, and for Arcturus $T_{\mathrm{eff}}=4280$ K, $\log
g=1.7$, $\xi_{\mathrm{micro}}=1.74$ km\,s$^{-1}$, [Fe/H]$=-0.53$,
[$\alpha$/Fe]$=+0.30$, see \cite{ryde_bulb2} for details.}.

These model atmospheres are  hydrostatic and are computed on the
assumptions of Local Thermodynamic Equilibrium (LTE), chemical
equilibrium, homogeneous plane-parallel (for the Sun) or
spherically-symmetric (for Arcturus) stratification, and the
conservation of the total flux (radiative plus convective; the
convective flux being computed using the local mixing length
recipe).

The synthetic spectra were calculated in plane parallel and
spherical symmetry for the Sun and Arcturus, respectively. We sample
the spectra with a resolution of  $R=600,000$. With a
micro-turbulence velocity of $1-2\,\mathrm{km\,s}^{-1}$, this will
ensure an adequate sampling. We finally convolve our synthetic
spectra, in order to fit the shapes and widths of the observed
lines, with a macro-turbulent (and instrumental) broadening,
represented by a radial-tangential function \cite{gray:1992}, with
$2.2$ and $3.7\,\mathrm{km\,s}^{-1}$ (FWHM), respectively.  The code
used for calculating the synthetic spectra is BSYN v. 7.09 which is
based on routines from the {\sc marcs} code.  A $^{12}\mathrm
C/^{13}\mathrm C$ ratio of $ 89$ is used for the Sun, and  of $9$
for Arcturus, see e.g. \cite{ryde_bulb2}.

The atomic line list used in our calculations is compiled from the
VALD database (\cite{vald}).  When needed, we determined
`astrophysical {$gf$}-values' by fitting atomic lines in the synthetic
spectra to the solar spectrum.  The lines fitted were, among others,
8 Fe, 3 C, 1 Ca, and 2 Ti lines.  In addition, the new strengths
($\log gf$) of the hfs Mn lines are calculated, but are given in a
relative scale. The lines fit the best when we scale the strengths
by a factor of 5.  The molecular line lists, which include CO, OH,
CN, SiO, CH,  were adopted as they are and instead of modifying the
{$gf$} values, the abundances of $\log\epsilon_{O}\mathrm{=8.63}$
(from OH lines), then $\log\epsilon_{C}\mathrm{=8.06}$ (from CO lines) and
last $\log\epsilon_{N}\mathrm{N=7.67}$ (from CN lines) were obtained
from the Arcturus atlas, in good agreement with \cite{ryde_bulb1}.
For the Sun a  $\log\epsilon_{CNO}\mathrm{=(8.41, 7.80, 8.66)}$
abundance is assumed.

\begin{figure}
\plotone{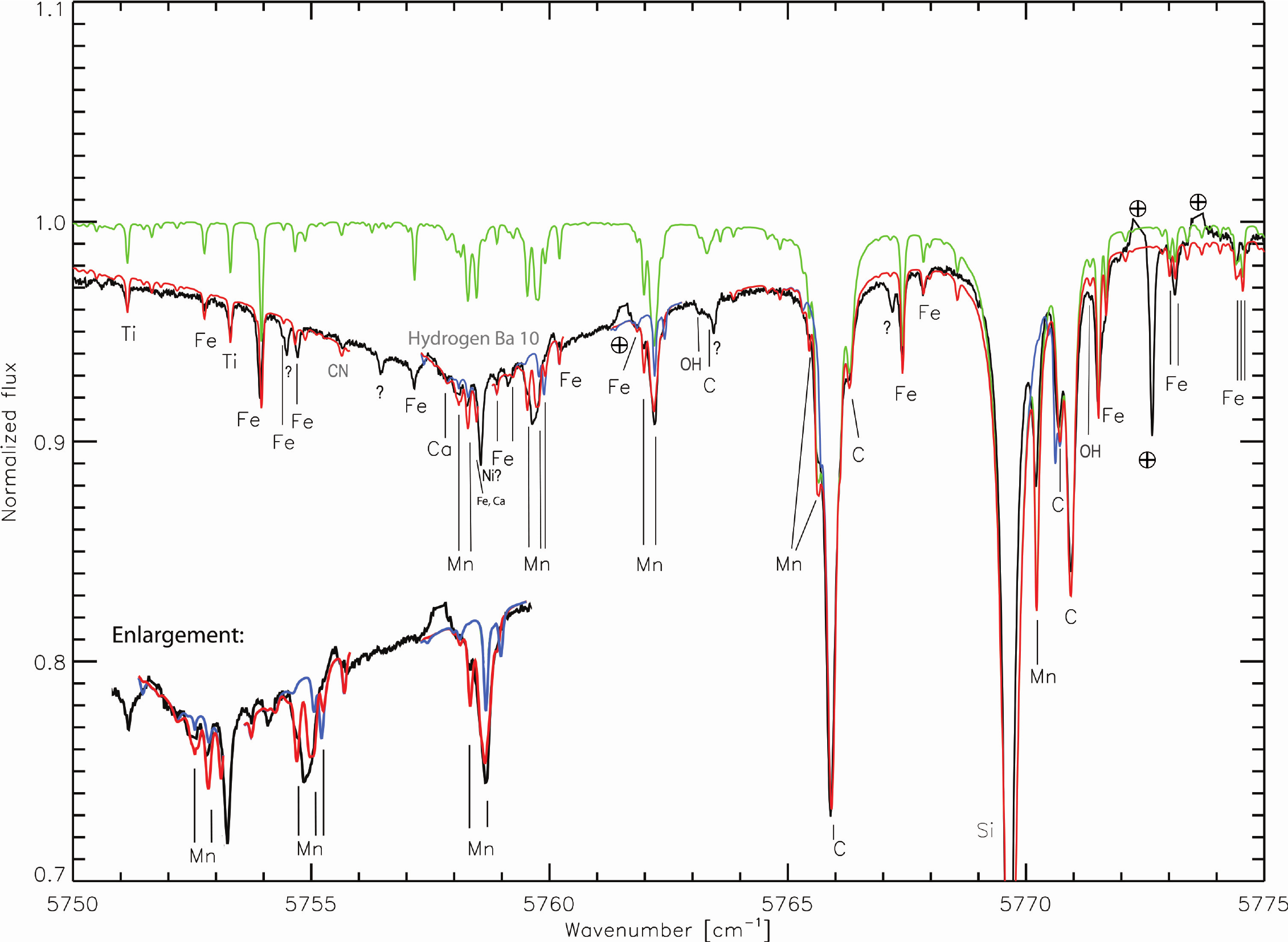}
\caption{\label{sun}
Section of the observed solar spectrum \cite{Livingston91} is shown in black. 
Our best synthetic spectra are shown in red, and include our newly
calculated Mn lines.  In green we show the same spectrum, but
omitting the Hydrogen Bracket 10 line, to show its influence. The
blue spectrum shows the spectrum using the Mn line list from the
VALD database.  All synthetic lines which are deeper than 0.97 of
the continuum are identified. A few features not identified are
labelled with question marks. Regions where the elimination of strong
telluric lines resulted in a degradation of the spectrum are marked
with an Earth symbol. In the lower left corner an enlargement of the spectrum
(not shown to scale) is plotted in order to show the fits in greater detail. }
\end{figure}

In Figure \ref{sun} we show our fits to the solar spectrum around the
Mn lines by plotting the normalized flux versus frequency given 
by the wavenumber in cm$^{-1}$. In this region the very wide Bracket 10 hydrogen 
line ($n=4-10$) dominates and complicates the comparison. Especially, the
normalization of the flux spectrum in this region is difficult since the continuum 
is absent over a wide frequency range due to the hydrogen line. Furthermore, 
existing codes calculating the broadening and strengths of solar hydrogen lines
cannot fit these lines. We have therefore manipulated the hydrogen
opacity by artificially changing the $\log gf$ value for this line
in order to fit the local "continuum" when analyzing the Mn lines.
The entire spectral region shown in Figure \ref{sun} is  more or
less affected by the hydrogen line. Thus, the original $\log
gf=-0.417$ is changed to $\log gf=-0.55$, except for the H-core
region, where for the low frequency side of the Ni line ($5756-5757$ 
cm$^{-1}$), it is changed to $\log gf=-0.75$ and on the high side ($5759-5760$ 
cm$^{-1}$), to $\log gf=-0.65$. The local synthetic spectra thus calculated are 
shown in the figure. We also show a synthetic spectrum with the original Mn line 
list (in blue) and compare this to the spectrum calculated with the new one
(in red). The new fit is very satisfactory. The Mn lines at $5770\,\mathrm 
{cm}^{-1}$ (17 325 \AA) are, however, too strong in the synthetic spectrum compared to the observed spectrum, the reason of which is not understood.

\begin{figure}
\plotone{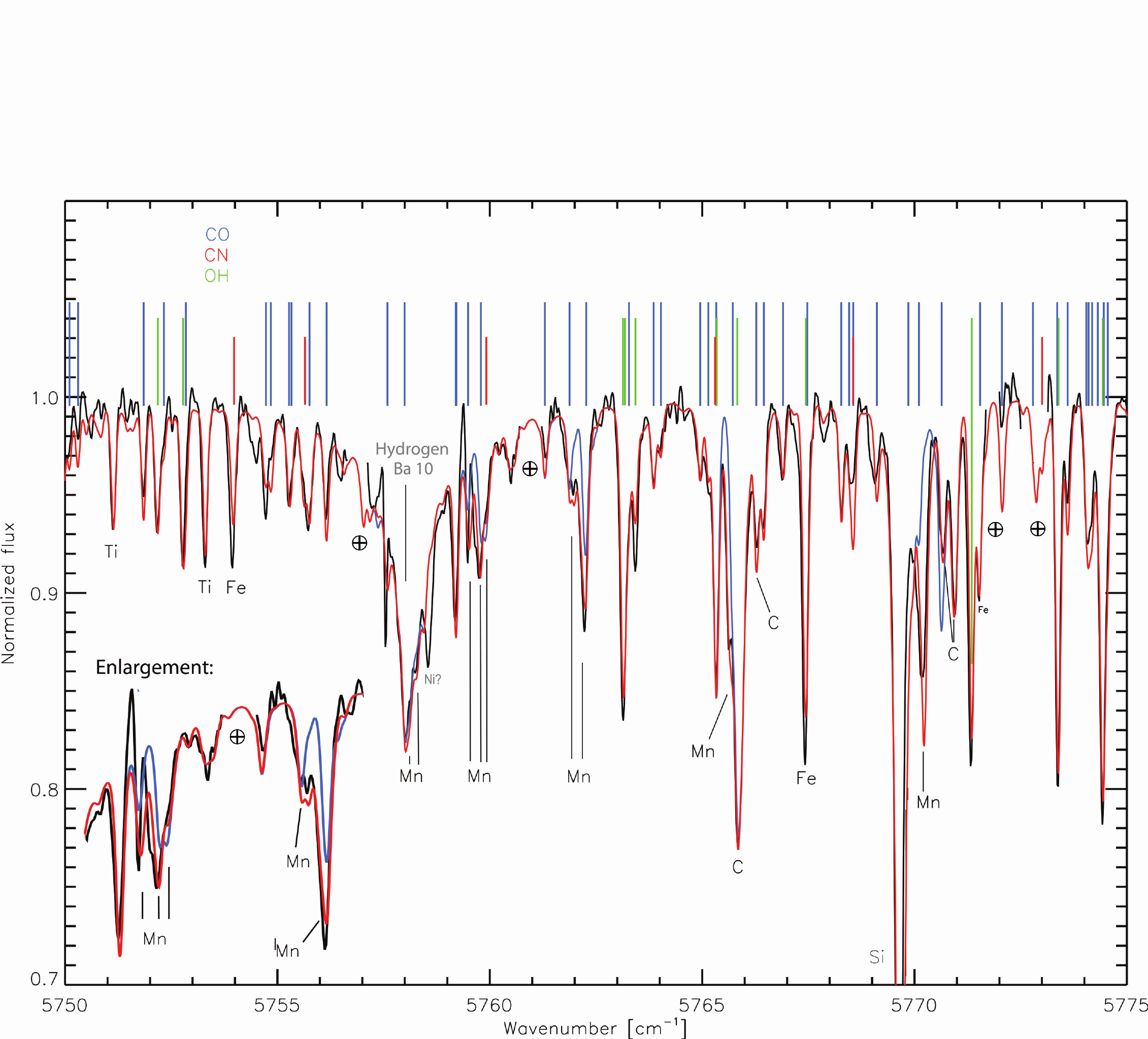} 
\caption{\label{aboo}
The wavelength region of interest for our newly calculated Mn lines is
displayed. The observed spectrum of the K1.5 III giant \emph{Arcturus}
(\cite{arcturusatlasII}) is shown in black and our best synthetic spectrum is
shown in red, which includes our newly calculated Mn lines. The blue
spectrum shows the spectrum using the Mn line list from the VALD
database.  All synthetic lines which are deeper than 0.97 of the
continuum are identified. Regions where the elimination of strong
telluric lines resulted in a degradation of the spectrum are marked
with an Earth symbol. In the lower left corner an enlargement of the spectrum
(not shown to scale) is plotted in order to show the fits in greater detail. }
\end{figure}

In Figure \ref{aboo} we present our synthetic fit to the observed
spectrum of the cooler giant star Arcturus. We see directly the
appearance of the many molecular lines from CO, CN, and OH, which
dominate the spectrum. The hydrogen line is now more narrow, as
expected for a lower gravity star, with less collisional broadening.
The hydrogen line was fitted by changing the opacity through a change
in the $\log gf_{\mathrm H Ba10}$  to $-0.7$. The general fit  to the
spectrum of this star is very good and especially the synthesized spectrum using the 
new data of the Mn lines (in red) is an improvement compared with the fit that was 
possible using previous data (in blue). Again, the Mn lines at 
$5770\,\mathrm {cm}^{-1}$ (17 325 \AA) are too strong in the synthesized spectrum, 
for reasons which require further investigations. We note however that these lines 
lie in the blue wing of a strong Si line, the broadening of which is not accurately 
synthesized.

As demonstrated in the figures of both the Sun and Arcturus, a few of the Mn lines are nearly absent in the spectra using the old data. For other Mn lines the residuals between the observed and synthesized spectra have more than halved when using the new data.

\section{Conclusions}
We have combined theoretical synthetic and experimental spectra of
the $3d^5 4s({^7S})4d$~e$^{6}$D -- $3d^5 4s({^7S})4f$~w$^{6}$F
17325, 17339, 17349, 17357 and 17362 \AA~lines in Mn~I
to derive information about the individual hyperfine lines that make
up these five spectral features. We have modelled the spectra of the sun and the red
giant star Arcturus using the new atomic data as well as using previously published atomic data and we have shown that our new data generates a better fit to observed stellar spectra. Using the new hyperfine structure data, these lines should therefore be useful in the analysis of stellar spectra.

Due to the extensive number of hyperfine transitions in this system we have 
concentrated our discussion on the three strong groups of hyperfine transitions 
called the 17325, 17339 and 17349 \AA~lines and only included the strongest of the 
transitions of each of these sub groups in the tables of this paper. A complete list 
of all the individual e$^6$D~--~w$^6$F hyperfine transitions can be found as online
material.

We have shown that the hyperfine levels involved in these transitions
can not be described in terms of the conventional hyperfine constants. 
Instead they have to be described individually due to the large
impact of the off-diagonal hyperfine interaction. By fitting our
theoretical spectra to experimental ones by allowing for small adjustments
to the calculated fine structure energies and hyperfine interaction
matrix elements in an iterative procedure we think we have developed
a method that could be applied to similar problems in other atomic
and ionic systems of interest to the astrophysical community.

\section{Acknowledgments}
Dr. Paul Barklem is thanked for discussions concerning the modelling of the
Hydrogen Ba 10 line and Dr. Kjell Eriksson for valuable help and
discussions concerning the running of the MARCS model-atmosphere
program.

M. Andersson is financed by the EU under the Science \& Technology
Fellowship Programme China (STF).

JG would like to thank the Nordic Centre at Fudan University, Shanghai, for supporting his visit to Fudan in 2013.

N. Ryde is a Royal Swedish Academy of Sciences Research Fellow
supported by a grant from the Knut and Alice Wallenberg Foundation.
N. Ryde also acknowledges support from the Swedish Research Council,
VR, and Funds from Kungl. Fysiografiska S\"allskapet i Lund.

EPSRC and PPARC of the UK supported the experimental studies undertaken 
by RWB while a PhD student at Imperial College London.

RH and YZ acknowledge the support of the National Natural Science Foundation of China under project no. 11074049 and by the Shanghai Leading Academic Discipline Project B107.

P. J\"onsson and T. Brage acknowledges support from the Swedish research
council.


\bibliographystyle{numbers}


\begin{deluxetable}{ccccclllc}
\tablecaption{\label{Tab-6D45}Line list of the
e$^6$D$_{9/2}$--w$^6$F transitions (the 17325 \AA~line) based on theoretical calculations. Only those 
with a $gf$-value of at least 10\% of the largest $gf$-value have been included. HFS is a hyperfine level index. See Section \ref{sec:results} for discussions about the wavenumber and $gf$-value uncertainties.}
\tablehead{ \multicolumn{2}{c}{w$^6$F} & \multicolumn{2}{c}{e$^6$D} & 
\multicolumn{2}{c}{Complete} &  & \multicolumn{1}{c}{Diagonal}  \\
\cline{5-6}  \cline{8-8} \colhead{{\sl F}} & \colhead{HFS} & \colhead{{\sl F}} & \colhead{HFS} &
\colhead{Wavenumber (cm$^{-1}$)$^\mathrm{a}$} & \colhead{{$gf^\mathrm{b}$}}
 & & \colhead{{$gf$}} &  \colhead{$\Delta${$gf$} (\%)} }
\startdata
  5 & 39 &  4 & 31 &  5770.2040 & 5.895 &&  6.047  &    3 \\
  4 & 38 &  3 & 30 &  5770.2033 & 4.875 &&  4.859  &   -0 \\
  3 & 37 &  2 & 29 &  5770.2047 & 3.999 &&  3.916  &   -2 \\
  6 & 42 &  5 & 32 &  5770.2126 & 7.193 &&  7.475  &    4 \\
  7 & 47 &  6 & 33 &  5770.2212 & 8.898 &&  9.153  &    3 \\
  8 & 54 &  7 & 34 &  5770.2474 & 1.109$\times 10^1$ &&  1.109$\times 10^1$  &    0 \\
  5 & 46 &  5 & 32 &  5770.2917 & 1.281 &&  9.455$\times 10^{-1}$ &  -26 \\
  6 & 52 &  6 & 33 &  5770.2983 & 1.707 &&  1.269  &  -26 \\
  7 & 61 &  7 & 34 &  5770.3259 & 2.219 &&  1.692  &  -24 \\
  \hline 
\multicolumn{9}{l}{$^\mathrm{a}$ \footnotesize Our recommended wavenumbers with an estimated overall relative uncertainty less than $0.02$ cm$^{-1}$.}  \\
\multicolumn{9}{l}{$^\mathrm{b}$ \footnotesize Our recommended $gf$-values with an estimated overall uncertainty of $5\%$.}
\enddata
\end{deluxetable}

\begin{deluxetable}{ccccclclc}
\tablecaption{\label{Tab-6D35}
Line list of the e$^6$D$_{7/2}$--w$^6$F transitions (the 17339 \AA~line) based on 
theoretical calculations. Only those with {$gf$}-values of at least 10\% of the 
largest {$gf$}-value have been included. HFS is a hyperfine level index. 
See Section \ref{sec:results} for discussions about the wavenumber~and~$gf$-value uncertainties.}

\tablehead{ \multicolumn{2}{c}{w$^6$F} & \multicolumn{2}{c}{e$^6$D}
& \multicolumn{2}{c}{Complete}
& & \multicolumn{1}{c}{Diagonal}  \\
 \cline{5-6}  \cline{8-8}
\colhead{{\sl F}} & \colhead{HFS} & \colhead{{\sl F}} & \colhead{HFS} &
\colhead{Wavenumber (cm$^{-1}$)$^\mathrm{a}$} & \colhead{{$gf^\mathrm{b}$}}
 & & \colhead{{$gf$}} &  \colhead{$\Delta${$gf$} (\%)} }
\startdata
  5 & 39 &  4 & 38 &  5765.4221 & 8.493$\times 10^{-1}$ & &    0                 & $-$ \\
  6 & 42 &  5 & 39 &  5765.4283 & 1.002                 & &    0                 & $-$ \\
  7 & 47 &  6 & 40 &  5765.4371 & 7.791$\times 10^{-1}$ & &    0                 & $-$ \\
  3 & 41 &  2 & 36 &  5765.5891 & 2.563                 & &2.704                 &    5 \\
  4 & 43 &  3 & 37 &  5765.5901 & 2.884                 & &3.650                 &   27 \\
  5 & 46 &  4 & 38 &  5765.5943 & 3.574                 & &4.831                 &   35 \\
  2 & 40 &  1 & 35 &  5765.6043 & 2.212                 & &1.991                 &  -10 \\
  6 & 52 &  5 & 39 &  5765.6137 & 4.899                 & &6.263                 &   28 \\
  4 & 50 &  4 & 38 &  5765.6496 & 1.207                 & &1.146                 &   -5 \\
  7 & 61 &  6 & 40 &  5765.6579 & 7.160                 & &7.964                 &   11 \\
  5 & 56 &  5 & 39 &  5765.6728 & 1.967                 & &1.753                 &  -11 \\
  2 & 45 &  2 & 36 &  5765.6911 & 7.873$\times 10^{-1}$ & &4.918$\times 10^{-1}$ &  -38 \\
  6 & 64 &  6 & 40 &  5765.7191 & 3.196                 & &2.595                 &  -19 \\
  3 & 55 &  3 & 37 &  5765.7455 & 7.959$\times 10^{-1}$ & &6.942$\times 10^{-2}$ &  -91 \\
  5 & 56 &  4 & 38 &  5765.7597 & 8.283$\times 10^{-1}$ & &4.023$\times 10^{-1}$ &  -51 \\
  4 & 58 &  4 & 38 &  5765.7982 & 9.405$\times 10^{-1}$ & &6.545$\times 10^{-2}$ &  -93 \\
  5 & 65 &  5 & 39 &  5765.8638 & 7.462$\times 10^{-1}$ & &4.363$\times 10^{-2}$ &  -94 \\
\hline 
\multicolumn{9}{l}{$^\mathrm{a}$ \footnotesize Our recommended wavenumbers with an estimated overall relative uncertainty less than $0.02$ cm$^{-1}$.}  \\
\multicolumn{9}{l}{$^\mathrm{b}$ \footnotesize Our recommended $gf$-values with an estimated overall uncertainty of $5\%$.} 
\enddata
\end{deluxetable}

\begin{deluxetable}{ccccclllc}
\tablecaption{\label{Tab-6D25}
Line list of the e$^6$D$_{5/2}$--w$^6$F transitions (the 17349 \AA~line) based on 
theoretical calculations. Only those with a {$gf$}-value of
at least 10\% of the largest {$gf$}-value have been included. HFS is a hyperfine 
level index. See Section \ref{sec:results} for discussions about the wavenumber and 
$gf$-value uncertainties.}

\tablehead{ \multicolumn{2}{c}{w$^6$F} & \multicolumn{2}{c}{e$^6$D}
& \multicolumn{2}{c}{Complete}
& & \multicolumn{1}{c}{Diagonal}  \\
\cline{5-6} \cline{8-8}
\colhead{{\sl F}} & \colhead{HFS} & \colhead{{\sl F}} & \colhead{HFS} &
\colhead{Wavenumber (cm$^{-1}$)$^\mathrm{a}$} & \colhead{{$gf^\mathrm{b}$}}
& & \colhead{{$gf$}} &  \colhead{$\Delta${$gf$} (\%)} }
\startdata
  5 & 46 &  4 & 45 &  5761.9605 & 1.632                 & & 0                     & $-$ \\
  4 & 43 &  3 & 44 &  5761.9661 & 1.428                 & & 0                     & $-$ \\
  6 & 52 &  5 & 46 &  5761.9707 & 1.265                 & & 0                     & $-$ \\
  3 & 41 &  2 & 43 &  5761.9790 & 8.594$\times 10^{-1}$ & & 0                     & $-$ \\
  3 & 48 &  2 & 43 &  5762.0822 & 8.391$\times 10^{-1}$ & & 1.720                 & 105 \\
  4 & 50 &  3 & 44 &  5762.0904 & 1.217                 & & 2.627                 & 116 \\
  2 & 45 &  1 & 42 &  5762.1138 & 1.319                 & & 1.032                 & -22 \\
  3 & 55 &  3 & 44 &  5762.1196 & 4.518$\times 10^{-1}$ & & 6.925$\times 10^{-1}$ &  53 \\
  5 & 56 &  4 & 45 &  5762.1261 & 2.124                 & & 3.783                 &  78 \\
  1 & 44 &  0 & 41 &  5762.1315 & 8.210$\times 10^{-1}$ & & 5.350$\times 10^{-1}$ & -35 \\
  4 & 58 &  4 & 45 &  5762.1431 & 1.309                 & & 1.484                 &  13 \\
  6 & 64 &  5 & 46 &  5762.1700 & 3.845                 & & 5.216                 &  36 \\
  3 & 55 &  2 & 43 &  5762.1840 & 7.301$\times 10^{-1}$ & & 6.677$\times 10^{-1}$ &  -9 \\
  2 & 53 &  2 & 43 &  5762.1971 & 9.588$\times 10^{-1}$ & & 2.473$\times 10^{-1}$ & -74 \\
  5 & 65 &  5 & 46 &  5762.2039 & 3.043                 & & 2.720                 & -11 \\
  4 & 58 &  3 & 44 &  5762.2413 & 6.580$\times 10^{-1}$ & & 6.595$\times 10^{-1}$ &   0 \\
  4 & 63 &  5 & 46 &  5762.2547 & 1.016                 & & 5.505$\times 10^{-1}$ & -46 \\
  3 & 57 &  3 & 44 &  5762.2582 & 1.126                 & & 0                     & $-$ \\
  2 & 60 &  3 & 44 &  5762.2783 & 6.592$\times 10^{-1}$ & & 1.802$\times 10^{-1}$ & -73 \\
  5 & 65 &  4 & 45 &  5762.2958 & 4.375$\times 10^{-1}$ & & 4.534$\times 10^{-1}$ &   4 \\
  4 & 63 &  4 & 45 &  5762.3559 & 8.882$\times 10^{-1}$ & & 1.126$\times 10^{-1}$ & -87 \\
  3 & 62 &  4 & 45 &  5762.3565 & 6.840$\times 10^{-1}$ & & 3.378$\times 10^{-1}$ & -51 \\
\hline 
\multicolumn{9}{l}{$^\mathrm{a}$ \footnotesize Our recommended wavenumbers with an estimated overall relative uncertainty less than $0.02$ cm$^{-1}$.}  \\
\multicolumn{9}{l}{$^\mathrm{b}$ \footnotesize Our recommended $gf$-values with an estimated overall uncertainty of $5\%$.}
\enddata
\end{deluxetable}

\end{document}